\documentclass[ reprint,
superscriptaddress,
twocolumn,
aps,
]{revtex4-2}
\usepackage{amssymb}
\usepackage{amsmath}
\usepackage{graphicx}
\usepackage{dcolumn}
\usepackage{bm}
\usepackage{xparse}

\NewDocumentCommand{\dint}{t\limits e{_^}}{  \mathop{
    \displaystyle \int
    \IfBooleanT{#1}{\limits}
    \IfValueT{#2}{_{#2}}
    \IfValueT{#3}{^{#3}}
  }}
\begin{document}

\title{Airy gas model: From three to reduced dimensions}
\author{K. Bencheikh}
\affiliation{Laboratory of Quantum Physics and Dynamical Systems. Department
of Physics. Ferhat Abbas University of Setif-1, Campus EI-Bez, Road of
Algiers, 19137 Setif, Algeria}
\author{A. Putaja}
\affiliation{Computational Physics Laboratory, Tampere University, 33101
Tampere, Finland}
\author{E. R{\"a}s{\"a}nen}
\affiliation{Computational Physics Laboratory, Tampere University, 33101
Tampere, Finland}
\date{\today}

\begin{abstract}
By using the propagator of linear potential as a main tool, we extend the Airy
gas model, originally developed for the three-dimensional ($d=3$) edge
electron gas, to systems in reduced dimensions ($d=2,1$). First, we derive
explicit expressions for the edge particle density and the corresponding
kinetic energy density (KED) of the Airy gas model in all dimensions. The
densities are shown to obey the local virial theorem. We obtain a functional
relationship between the positive KED and the particle density and its
gradients and analyze the results inside the bulk as a limit of the
local-density approximation. We show that in this limit the KED functional
reduces to that of the Thomas-Fermi model in $d$ dimensions.
\end{abstract}

\maketitle

\section{\label{sec1}Introduction}

The Thomas-Fermi (TF) theory~\cite{Dreizler-Gross} is one of the first
approaches towards the widely used density-functional theory (DFT)
\cite{HK-64}. Both theories are built on the central role of the particle
density in the study of many-particle systems. The TF model gives the exact
kinetic energy of the uniform electron gas, as well as the correct
ground-state energy asymptotics for large atomic
numbers~\cite{Lieb-73,Lieb-81}. For finite $N$, however, the TF model becomes
a crude approximation; for example, it predicts unstable negatively charged
ions and does not describe atomic binding at all.

The TF model has been improved by the inclusion of inhomogeneity corrections
through a gradient expansion for the kinetic energy and the
exchange-correlation functionals~\cite{Englert-88}. A significant improvement
was the developement of the so-called generalized gradient approximation (GGA)
\cite{Karasiev-2012}, which was followed by more accurate functionals such as
the meta-GGAs \cite{Eich-2014}. An alternative correction to the TF model was
recently developed by Ribeiro et al.~\cite{Ribeiro-2015,Ribeiro-2017} based on
the use of a uniform semiclassical approximation. These leading corrections to
the TF model substantially improve the description of the pointwise particle
and the kinetic energy densities (KEDs) in one dimension (1d) without any
gradient expansion~\cite{Ribeiro-2018}. However, further generalizations to
higher dimensions are called for. Along this path, we mention a recent study
dealing with systematic corrections to the TF model in three dimensions (3d)
without a gradient expansion through the use of the unitary evolution
operator~\cite{Englert2018}. That work focuses on the so-called
potential-functional theory, which employs the single-particle potential on an
equal footing with the density~\cite{Elliot-2015}.

In a landmark work, Kohn and Mattsson~\cite{KohnMattson} introduced the concept
of the \emph{edge electron gas} as a convenient way to deal with physical
systems having edge regions. The resulting theoretical treatment is known as
the \emph{Airy gas model}, which adapts to the changes in the particle density
from the bulk behavior to evanescence. The simplicity of the Airy gas model
lies in the fact that the effective potential near the edges is approximated
by a linear potential. Consequently, the normalized single-particle wave
functions, e.g., in the Kohn-Sham picture, are proportional to the Airy
function. As a result, the Airy gas model constitutes an important improvement
of both the TF theory and DFT when describing these regions at jellium
surfaces, for example. The model has inspired the development of density
functionals within DFT. For instance the Airy gas model has been used to
construct an exchange-energy functional, and test calculations prove to be
better than the generalized gradient approximation~\cite{vitos1}. Moreover the
designed AM05 functional~\cite{armiento1}, is an exchange-correlation
functional tailored for an accurate treatment of systems with electronic
surfaces and has excellent performance also for solids~\cite{armiento2}.

Here we derive explicit expressions for the edge particle density and for the
corresponding edge KED in all spatial dimensions ($d=3,2,1$). We use the
propagator of the linear potential as the main tool, for which explicit
analytical expressions exist in all dimensions. This approach has the
advantage to avoid the explicit use of wave functions. In particular, the
particle density and the KED are given as appropriate inverse Laplace
transforms of the Bloch propagator.

Our paper is organized as follows. In Sec. II we obtain the Bloch propagator
associated with the Airy gas model. Then we employ the propagator in Sec. III to
obtain explicit analytical expressions for the particle densities
in all dimensions $d=3,2,1$. In Sec. IV we continue the procedure to obtain explicit
expressions for the KEDs in all dimensions, including also the expressions
for the so-called kinetic energy refinement or enhancement factor
defined as the ratio of kinetic energy density relative to that of the TF
theory. In the $d=3$ case our results are compared with those obtained 
earlier by Vitos~\cite{vitos1}. In Sec. V we show that the
derived densities and KEDs obey the so-called local virial theorem. 
Finally, in Sec. VI we analyze the limit of the local-density approximation
(LDA) of Airy gas model inside the bulk. In particular, we show how in this
limit our KED functional reduces to that of the TF model in $d$ dimensions.
The paper ends with a brief summary in Sec. VII.

\section{Bloch propagator}

In the following we derive an analytical closed form of the so-called Bloch
propagator associated with the Airy gas model. The main advantage in using a
propagator approach is the fact that no explicit use of occupied
single-particle states is required. Moreover, as we will see the use of a
propagator as a tool allows us to deal with a unified description in all the dimensions.

Let us consider a system of $N$ independent fermions moving in some known
potential $V(\bar{r})$. The one-body density matrix in
zero temperature can be written by means of the unit-step function
$\theta(x)$ as follows
\begin{equation}
\rho(\bar{r},\bar{r}^{\prime};\mu)=\sum_{n}\phi
_{n}(\bar{r})\phi_{n}^{\ast}(\bar{r}^{\prime})\theta
(\mu-\varepsilon_{n}), \label{eq1}%
\end{equation}
where the sum is computed over occupied single-particle states up to the Fermi
energy $\mu$. The single-particle wave functions $\phi_{n}$ are the normalized
solutions of the Schr\"{o}dinger equation $\hat{H}\phi_{n}=\varepsilon_{n}\phi_{n}$
with the Hamiltonian
\begin{equation}
\hat{H}=-\frac{\hbar^{2}}{2m}\nabla^{2}+V(\bar{r}),
\label{eq2}%
\end{equation}
and $\varepsilon_{n}$ are the single-particle energies. The unit-step function
can be written as
\begin{equation}
\theta(\mu-\varepsilon_{n})=%
{\displaystyle \int \limits_{c-i\infty}^{c+i\infty}}
\frac{d\eta}{2\pi i}\frac{e^{\eta(\mu-\varepsilon_{n})}}{\eta} \label{eq3}%
\end{equation}
with $c>0$ \cite{Abramowitz-Stegun}. This allows us to write the density
matrix in Eq. (\ref{eq1}) as \cite{Brack-Bhaduri}
\begin{equation}
\rho(\bar{r},\bar{r}^{\prime};\mu)=%
{\displaystyle \int \limits_{c-i\infty}^{c+i\infty}}
\frac{d\eta}{2\pi i}e^{\eta \mu}\frac{U(\bar{r},\bar
{r}^{\prime};\eta)}{\eta}. \label{eq4}%
\end{equation}
Here $U(\bar{r},\bar{r}^{\prime};\eta)$ is the matrix
element of the Bloch operator $\hat{U}=e^{-\eta \hat{H}}$, i.e.,
\begin{equation}
U(\bar{r},\bar{r}^{\prime};\eta)=\sum_{n}\phi
_{n}(\bar{r})\phi_{n}^{\ast}(\bar{r}^{\prime})\exp
(-\eta \varepsilon_{n}). \label{eq5}%
\end{equation}
Depending on the nature of the parameter $\eta$, the above quantity is
referred as a heat kernel, canonical Bloch density, or time evolution
propagator~\cite{Dreizler-Gross}. Here $\eta$\ is defined as a complex
variable, and we shall call $U(\bar{r},\bar{r}^{\prime
};\eta)$ as the Bloch propagator.

Let us consider the following one-particle Hamiltonian:
\begin{align}
\hat{H} &  =\left[  -\frac{\hbar^{2}}{2m}\left(  \frac{\partial^{2}}{\partial x^{2}%
}+\frac{\partial^{2}}{\partial y^{2}}\right)  -\frac{\hbar^{2}}{2m}%
\frac{\partial^{2}}{\partial z^{2}}+Fz\right]  \text{, }\nonumber \\
&  -\frac{L_{x}}{2}\leq x\leq \frac{L_{x}}{2}\text{ ,}-\frac{L_{y}}{2}\leq
y\leq \frac{L_{y}}{2}\text{, }-\infty<z<+\infty \label{eq6}\\
\hat{H} &  =0,\text{ \  \  \  \  \  \  \  \  \  \  \  \  \  \  \  \  \  \  \  \  \  \  \  \ elsewhere.}%
\label{eq7}%
\end{align}
This Hamiltonian describes a particle with mass $m$ subjected to a constant
potential inside a cross-sectional area $A=L_{x}L_{y}$ in two dimensions
$(x,y)$, and to a linear potential in the third direction $z$. The
corresponding propagator $\hat{U}=e^{-\eta \hat{H}}$, can be factorized as a
product, $\hat{U}^{d=3}=\hat{U}_{\eta}^{x}\hat{U}_{\eta}^{y}\hat{U}_{\eta}^{z}$. When the lengths
$L_{x}$ and $L_{y}$ are expected to approach infinity as assumed in the Airy
gas model, $\hat{U}_{\eta}^{x}$ and $\hat{U}_{\eta}^{y}$ can be taken to be the
free-particle propagators along $x$ and $y$ directions,
respectively~\cite{Feynman-Hibbs}. That is,
\begin{align}
U_{\eta}^{x}(x,x^{\prime}) &  =\left(  \frac{m}{2\pi \hbar^{2}\eta}\right)
^{\frac{1}{2}}\exp \left[  -\frac{m}{2\hbar^{2}\eta}(x-x^{\prime}%
)^{2}\right]  \label{eq8}\\
U_{\eta}^{y}(y,y^{\prime}) &  =\left(  \frac{m}{2\pi \hbar^{2}\eta}\right)
^{\frac{1}{2}}\exp \left[  -\frac{m}{2\hbar^{2}\eta}(y-y^{\prime}%
)^{2}\right]  .\label{eq9}%
\end{align}
The propagator for the linear potential along the $z$ direction is exactly
known~\cite{Feynman-Hibbs} and has the form
\begin{align}
U_{\eta}^{z}(z,z^{\prime}) &  =\left(  \frac{m}{2\pi \hbar^{2}\eta}\right)
^{\frac{1}{2}}\exp \left(  \frac{\hbar^{2}}{24m}\eta^{3}F^{2}\right)
\nonumber \\
\times &  \exp \left[  -\eta F\left(  \frac{z+z^{\prime}}{2}\right)  \right]
\times \exp \left[  -\frac{m}{2\hbar^{2}\eta}(z-z^{\prime})^{2}\right]
.\label{eq10}%
\end{align}
Since $\hat{U}^{d=3}=\hat{U}_{\eta}^{x}\hat{U}_{\eta}^{y}\hat{U}_{\eta}^{z}$ , we then obtain the Bloch
propagator for the Airy gas model in $d$ dimensions
\begin{align}
U^{(d)} &  (\bar{r},\bar{r}^{\prime};\eta)=\left(
\frac{m}{2\pi \hbar^{2}\eta}\right)  ^{\frac{d}{2}}\exp \left(  \frac{\hbar^{2}%
}{24m}\eta^{3}F^{2}\right)  \nonumber \\
\times &  \exp \left[  -\eta F\left(  \frac{z+z^{\prime}}{2}\right)  \right]
\nonumber \\
\times &  \exp \left[  -\frac{m}{2\hbar^{2}\eta}(  \bar{r}%
-\bar{r}^{\prime})^{2}  \right],
\label{eq11}%
\end{align}
where $\bar{r}$ and\  $\bar{r}^{\prime}$ are
$d$-dimensional position vectors.
Here for $U$ (and below for $\rho$ and $\tau$) we denote the dimension $d$ in parentheses in the superscript. For $d=3$, Eq.~(\ref{eq11}) can 
be interpreted as the Bloch propagator associated with the
Hamiltonian of Eq. (\ref{eq6}) in the limits $L_{x}\rightarrow \infty$ and
$L_{y}\rightarrow \infty$. In a similar way, the resulting propagator for $d=2$ is associated with the two-dimensional version of the Hamiltonian in 
Eqs. (\ref{eq6})-(\ref{eq7}). It describes the motion of the particles in the
$xz$ plane, where free motion is assumed along the $x$-axis. For $d=1$, the particles are assumed to move only along the $z$-axis and subjected to a linear potential.

In the following, we show that the Bloch propagator of Eq. (\ref{eq11}) is
associated to the Hamiltonian of the Airy gas model of Kohn and Mattsson in
$d=3$. The Hamiltonian of this model
reads\cite{KohnMattson,Lind-Matt-Armiento}
\begin{align}
\hat{H}  &  =\left[  -\frac{\hbar^{2}}{2m}\left(  \frac{\partial^{2}}{\partial
x^{2}}+\frac{\partial^{2}}{\partial y^{2}}\right)  -\frac{\hbar^{2}}{2m}%
\frac{\partial^{2}}{\partial z^{2}}+v_{\mathrm{eff}}(z)\right]  \text{,
}\nonumber \\
&  -\frac{L_{x}}{2}\leq x\leq \frac{L_{x}}{2}\text{, }-\frac{L_{y}}{2}\leq
y\leq \frac{L_{y}}{2}\text{, }-L<z<+\infty \label{eq12}\\
{H}  &  =0\text{ \  \  \  \  \  \  \  \  \  \  \  \  \  \  \  \  \  \  \  \  \  \  \  \  \  \ elsewhere}.
\label{eq13}%
\end{align}
Here $v_{\mathrm{eff}}(z)$ is the confining potential along the $z$ direction
given by
\begin{align}
v_{\mathrm{eff}}(z)  &  =\infty,\text{ \  \  \ }z\leq-L\label{eq14}\\
v_{\mathrm{eff}}(z)  &  =Fz,\text{ \  \  \ }z>-L, \label{eq15}%
\end{align}
where $F=dv_{\mathrm{eff}}(z)/dz$ \ is the slope of the effective potential.
The characteristic length scale is given by $l=\left(  \hbar^{2}/(2mF)\right)
^{1/3}$ with the corresponding energy $\widetilde{\varepsilon}=Fl=\left(
\hbar^{2}F^{2}/(2m)\right)  ^{1/3}$. The normalized eigenfunctions $\psi_{r}$
with eigenvalues $E_{r}$ of the KS equations are of the
form~\cite{KohnMattson}
\begin{equation}
\psi_{r}(x,y,z)=\frac{1}{\sqrt{L_{x}L_{y}}}e^{\frac{i}{\hbar}p_{x}x}%
e^{\frac{i}{\hbar}p_{y}y}\phi_{j}(z) \label{eq16}%
\end{equation}
with $r\equiv(j,p_{x},p_{y})$ and $\ p_{i}L_{i}=2\pi \hbar m_{i}(i=x,y)$, and
$A\equiv L_{x}L_{y}$ is the cross-sectional area. The functions $\phi_{j}(z)$
obey%
\begin{equation}
\left[  -\frac{\hbar^{2}}{2m}\frac{d^{2}}{dz^{2}}+Fz\right]  \phi
_{j}(z)=\varepsilon_{j}\phi_{j}(z). \label{eq17}%
\end{equation}
and the occupied states have energies $E_{r}$ so that
$E_{r}=(p_{x}^{2}+p_{y}^{2})/2m+\varepsilon_{j}\leq \mu$. In the Airy gas
model, following the arguments of Ref.~\cite{KohnMattson} in the limit
$L\rightarrow \infty$, the eigenvalues form a continuous spectrum. Therefore,
in this limit the Hamiltonian in Eq. (\ref{eq12}) becomes compatible with the
one given in Eq. (\ref{eq6}). Therefore, we can consider the Bloch propagator
$U(\bar{r},\bar{r}^{\prime};\eta)$ found in
Eq.~(\ref{eq11}) for the system under consideration. Furthermore, we use an
absolute energy scale as was done in previous works on Airy gas
model~\cite{KohnMattson,Lind-Matt-Armiento}, so that the Fermi energy
is set to zero, i.e., $\mu=0$. With this choice, the expression of the diagonal Bloch propagator in $d$ dimensions reduces to
\begin{align}
U^{(d)}(\bar{r},\bar{r};\eta)= &  \left(  \frac{m}%
{2\pi \hbar^{2}\eta}\right)  ^{\frac{d}{2}}
\exp\left(\frac{\hbar^{2}}%
{24m}\eta^{3}F^{2}-\eta Fz\right).\label{eq18a}%
\end{align}

\section{Particle density in $d$ dimensions}

Here we utilize the Bloch propagator to derive explicit expressions
for the particle density in $d$ dimensions. The result for the $d=3$ case can be
compared to the alternative derivation reported in Refs. \cite{Lind-Matt-Armiento} and
\cite{Dean-Doussal-Majumdar-Schehr}. The results for the particle densities in
reduced dimensions ($d=2,1$) have particular relevance for applications
in low-dimensional systems such as quantum wells and wires.

The particle density for the Airy gas model can be obtained from the Bloch propagator in Eq.~(\ref{eq18a}) as
\begin{align}
\rho^{(d)}(z) 
& =\dint \limits_{c-i\infty }^{c+i\infty }\frac{d\eta }{2\pi i}%
\frac{U^{(d)}(\bar{r},\bar{r};\eta )}{\eta }  \label{eq18b} \\
&=\left(\frac{m}{2\pi \hbar^{2}}\right)^{\frac{d}{2}}%
{\displaystyle \int \limits_{c-i\infty}^{c+i\infty}}
\frac{d\eta}{2\pi i}\frac{e^{\frac{\hbar^{2}}{24m}\eta^{3}F^{2}-\eta Fz}}%
{\eta^{1+d/2}}. \label{eq18}%
\end{align}

To evaluate the integral representation of the density in  Eq. (\ref{eq18}), we
will first use the identity
\begin{equation}
\frac{1}{\eta^{1+d/2}}=\frac{1}{\Gamma \left(  1+\frac{d}{2}\right)  }%
{\displaystyle \int \limits_{0}^{\infty}}
e^{-\eta q}q^{\frac{d}{2}}dq \label{eq19}%
\end{equation}
 secondly, we change the variables $u=2^{-2/3}\widetilde{\varepsilon}\eta$, 
$\xi=z/l$, $v=2^{2/3}q/\widetilde{\varepsilon},$
and finally, using the integral representation of the Airy function \cite{Vallee-Soares}
\begin{equation}
A_{i}(t)=%
{\displaystyle \int \limits_{c-i\infty}^{c+i\infty}}
\frac{du}{2\pi i}\exp \left[  \frac{u^{3}}{3}-ut\right]  \label{eq22}%
\end{equation}
we obtain the density of the Airy gas (AG) model in the
$d$-dimensional form
\begin{align}
\rho^{(d)}(z)&=D_{d}\,g_{s}
{\displaystyle \int \limits_{0}^{\infty}}
dv\,v^{d/2}A_{i}\left(  2^{2/3}\xi+v\right),\label{eq23a}
\end{align}
where
\begin{align}
D_{d}(z)=\frac{2^{-d/3}}{(4\pi l^{2})^{d/2}}\frac{1}{\Gamma(1+d/2)}.
\label{eq23b}
\end{align}
where we have included a factor $g_{s}$ to account for the spin degeneracy. 

Applying Eq. (\ref{eq23a}) to $d=1,2,3$ and using the properties of Airy function 
in Eqs. (\ref{a3}), (\ref{a4}) and (\ref{a6}) in Appendix \ref{appendixA},  we obtain 
\begin{align}
\rho^{d=1}(z)&=g_{s}\frac{1}{l}\left[  A_{i}^{\prime2}\left(  \xi \right)  -\xi
A_{i}^{2}\left(  \xi \right)  \right]  . \label{eq38}\\
\rho^{d=2}(\xi)&=-g_{s}\frac{1}{4\pi l^{2}}\left[  2^{-2/3}A_{i}^{\prime}\left(
2^{2/3}\xi \right)  +\xi A_{i1}\left(  2^{2/3}\xi \right)  \right],
\label{eq33}\\
\rho^{d=3}(z)&=g_{s}\frac{1}{12\pi l^{3}}\left[  2\xi^{2}A_{i}^{2}\left(
\xi \right)  -A_{i}\left(  \xi \right)  A_{i}^{\prime}\left(  \xi \right)  -2\xi
A_{i}^{\prime2}\left(  \xi \right)  \right], \label{eq26}
\end{align}
where in the $d=2$ case of Eq.~(\ref{eq33}) we have
\begin{equation}
A_{i1}\left(  t\right)  =%
{\displaystyle \int \limits_{t}^{\infty}}
A_{i}\left(  v\right) dv. \label{eq32}%
\end{equation}

For an unpolarized system of fermions we have $g_{s}=2$. Thus, Eq.
(\ref{eq26}) leads to an expression that is identical to the one derived in
Refs. \cite{Lind-Matt-Armiento} and \cite{Dean-Doussal-Majumdar-Schehr}.

We remind that the above expressions for the particle density
were obtained by using a propagator of the linear potential adapted to the
Airy gas model. Hence, the results were obtained \emph{without} explicitly using the 
set of occupied single particle wave functions, in contrast with the
$d=3$ result in Ref. \cite{Lind-Matt-Armiento}. Recently, the densities have been 
found through n-point correlation functions of free fermions in a $d$-dimensional 
trap \cite{Dean-Doussal-Majumdar-Schehr}.


\section{Kinetic-energy density}

\subsection{Generic expressions in $d$ dimensions}

Motivated by the development of density functionals, the objective
in this section is to obtain a relationship between the positive KED and the
particle density and its gradients for arbitrary dimension $d=1,2,3$ in the
Airy gas model.
In the Kohn-Sham version of DFT \cite{KS65}, the interacting system is mapped
to non-interacting one of independent fermions. As a consequence, the total
noninteracting kinetic energy, $T[\rho]=%
{\textstyle \int}
\tau_{G}(\bar{r})d\bar{r}$, as any other observable, is
a functional of $\rho$. An \emph{explicit} density functional of a
noninteracting KED corresponds to an orbital-free DFT without the need for the
calculation of single-particle wave functions.

In the literature three different formulations for the KED are considered in
terms of the single-particle wave
functions~\cite{Lombard-Mas-Moszkowski,Brack2003}. The {\em Laplacian} (L) form is given
by
\begin{equation}
\tau_{L}(\bar{r})=-\frac{\hbar^{2}}{2m}\sum \limits_{n}\left[
\phi_{n}^{\ast}(\bar{r})\nabla^{2}\phi
_{n}(\bar{r})\right]  \theta(\mu-\varepsilon_{n}). \label{eq39}%
\end{equation}
This form obtained from the Schr\"{o}dinger equation
and can locally take positive or negative values. On the
other hand, the positively defined {\em gradient} (G) form of the KED, which is
generally considered in the Kohn-Sham version of DFT~\cite{KS65}, reads
\begin{equation}
\text{\  \ }\tau_{G}(\bar{r})=\frac{\hbar^{2}}{2m}\sum
\limits_{n}\left \vert \nabla\phi_{n}(\bar
{r})\right \vert ^{2}\theta(\mu-\varepsilon_{n}). \label{eq40}%
\end{equation}
Finally, we can consider the arithmetic mean of the Laplacian and gradient forms, i.e.,
i.e.,
\begin{equation}
\tau(\bar{r})=\left[  \tau_{L}(\bar{r})+\tau
_{G}(\bar{r})\right]  /2. \label{eq41}%
\end{equation}
We point out that while all these three expressions $\tau
_{L}(\bar{r})$, $\tau_{G}(\bar{r})$ and $\tau
(\bar{r})$ differ locally, they yield the same
total kinetic energy when integrated over the spatial coordinates.
For a spin-unpolarized system we can show that \cite{Brack2003}
\begin{equation}
\tau_{L}(\bar{r})=\tau_{G}(\bar{r})-\frac{\hbar^{2}}%
{4m}\nabla^{2}\rho(\bar{r}), \label{eq42}%
\end{equation}
where $\rho(\bar{r})=\sum_{n}\left \vert \phi_{n}(\bar
{r})\right \vert ^{2}\theta(\mu-\varepsilon_{n})$ is the diagonal part of the
density matrix in Eq. (\ref{eq1}). Combining the two previous
expressions yields
\begin{equation}
\tau_{G}(\bar{r})=\tau(\bar{r})+\frac{\hbar^{2}}%
{8m}\nabla^{2}\rho(\bar{r}). \label{eq43}%
\end{equation}

In the subsequent analysis, it turns out to be more convenient to
first use the mean KED $\tau(\bar{r})$,\ which can be
expressed in terms of the density matrix as \cite{Brack2003}
\begin{equation}
\tau(\bar{r})=-\frac{\hbar^{2}}{2m}\left[ \nabla_{\bar{s}}^{2}\rho \left(  \bar{R}+\frac{\bar
{s}}{2},\bar{R}-\frac{\bar{s}}{2};\mu \right)  \right]
_{\bar{s}=\bar{0},\bar{R}=\bar{r}}.
\label{eq44}%
\end{equation}
Here $\bar{R}=(\bar{r}+\bar{r}^{\prime})/2$ and
$\bar{s}=\bar{r}-\bar{r}^{\prime}$ denote the
centre-of-mass and relative coordinates, respectively. Inserting Eq.
(\ref{eq4}) into Eq. (\ref{eq44}) yields
\begin{align}
\tau(\bar{r})  &  =-\frac{\hbar^{2}}{2m}%
{\displaystyle \int \limits_{c-i\infty}^{c+i\infty}}
\frac{d\eta}{2\pi i}\frac{e^{\eta \mu}}{\eta}\nonumber \\
&  \times \left[  \nabla_{\bar{s}}^{2}U\left(
\bar{R}+\frac{\bar{s}}{2},\bar{R}%
-\frac{\bar{s}}{2};\eta \right)  \right]  _{\bar
{s}=\bar{0},\bar{R}=\bar{r}}. \label{eq45}%
\end{align}

The Laplace operator targeting the last row of Eq. (\ref{eq11}). Since $\nabla_{\bar{s}}\cdot\bar{s}=d$, it is easy to deduce
$\nabla_{\bar{s}}^{2}[\exp(-\frac{m}{2\hbar^{2}\eta
}\bar{s}^{2})]=-md/(\hbar^{2}\eta)$ \ for $\bar{s}%
=\bar{0}$. With this latter result, the mean KED in Eq. (\ref{eq45})
of the Airy gas in $d$ dimensions becomes

\begin{equation}
\tau^{(d)}(\bar{r})=\frac{d}{2}%
{\displaystyle \int \limits_{c-i\infty}^{c+i\infty}}
\frac{d\eta}{2\pi i}\frac{U^{(d)}(\bar{r},\bar{r};\eta
)}{\eta^{2}},\label{eq47}%
\end{equation}
where as -- previously mentioned -- we take $\mu=0$.

Next, let us insert Eq. (\ref{eq18a}) into Eq. (\ref{eq47}) and after that use 
Eqs. (\ref{eq18}) and (\ref{eq23a}) to obtain
\begin{align}
\tau^{(d)}(z)&=\frac{d}{2}\left(  \frac{m}{2\pi \hbar^{2}}\right)  ^{d/2}%
{\displaystyle \int \limits_{c-i\infty}^{c+i\infty}}
\frac{d\eta}{2\pi i}\frac{e^{\frac{\hbar^{2}}{24m}\eta^{3}F^{2}-\eta Fz}}%
{\eta^{2+d/2}} \label{eq49a} \\
&= \frac{d}{2}\frac{2\pi \hbar^{2}}{m}\rho^{(d+2)}(z) \\
&= \frac{\hbar^{2}}{2m}\frac{d}{d+2}\frac{2^{-2/3} }{ l^{2}} D_{d}
 {\displaystyle \int \limits_{0}^{\infty}}
dv\,v^{1+d/2}A_{i}\left(  2^{2/3}\xi+v\right). \label{eq50a}
\end{align}

The second derivative $\rho^{(d)}$ with respect to $\xi $ leads to 
\begin{align}
\frac{\partial^{2}\rho^{(d)}(\xi)}{\partial \xi^{2}}
&=D_{d} 2^{4/3}
{\displaystyle \int \limits_{0}^{\infty}}
dv\,v^{d/2}A^{\prime \prime}_{i}\left(  2^{2/3}\xi+v\right)\nonumber \\
&= 4\xi\rho^{(d)}(\xi) \nonumber \\
&+ 2^{4/3} D_{d}
{\displaystyle \int \limits_{0}^{\infty}}
dv\,v^{1+d/2}A_{i}\left(  2^{2/3}\xi+v\right), \label{der2rho}
\end{align}
where in the second line we have used Eq. (\ref{a1}). In this expression the integral of the last term is same as Eq. (\ref{eq50a}). So the mean KED in $d$ dimensions can be written as
\begin{align}
\tau^{(d)}(z)
= \frac{\hbar^{2}}{2m}\frac{d}{d+2}\frac{1}{4 l^{2}}
\left(\frac{\partial^{2}\rho(\xi)}{\partial \xi^{2}} - 4\xi\rho(\xi)\right). \label{mean_ked}
\end{align}

In DFT the positive KED defined in Eq. (\ref{eq40}) is used when developing
approximate KED functionals. We will use Eq. (\ref{eq43}) and Eq. (\ref{mean_ked}) to obtain the expression of the positively defined KED in the gradient form as
\begin{align}
\tau^{(d)}_{G}(\bar{r})&=\tau^{(d)}(\bar{r})+\frac{\hbar^{2}}%
{8m}\nabla^{2}\rho^{(d)}(\bar{r})\nonumber \\
&= \frac{\hbar^{2}}{2m}\frac{d}{d+2}\frac{1}{4 l^{2}}
\left(\rho^{\prime\prime}(\xi) - 4\xi\rho(\xi)\right)
+\frac{\hbar^{2}}%
{8m l^{2}}\frac{\partial^{2}}{\partial \xi^{2}}\rho(\xi)\nonumber \\
&=  \frac{\hbar^{2}}{2m l^{2}}\left[- \frac{d}{d+2}\xi \rho(\xi)+
\frac{1}{2}\frac{d+1}{d+2}\rho^{\prime\prime}(\xi)\right] \label{pos_ked_a}
\end{align}

To obtain a KED\ functional of the density $\rho$, it remains
to eliminate variable $\xi=z/l$ from Eq. (\ref{pos_ked_a}). Therefore, we may express $\xi$  in terms of the particle density and its
derivatives. Here we focus on the main result and leave the details of the derivation in Appendix \ref{appendixB}, where we find $\xi$ in a $d$%
-dependent form as
\begin{equation}
\xi=\frac{d}{2}\frac{\rho}{\rho^{\prime}}+\frac{\rho^{\prime \prime \prime}%
}{4\rho^{\prime}},\text{ \  \  \  \ }d=1,2,3.\label{eq71a}%
\end{equation}
Substituting this result into the expression of the positive KED in Eq.
(\ref{pos_ked_a}) leads to a density functional
\begin{align}
\tau_{G}^{(d)}\left[  \rho \right]  = &  \frac{\hbar^{2}}{2ml^{2}}\left[
-\frac{d}{d+2}\left(  \frac{d}{2}\frac{\rho}{\rho^{\prime}}+\frac{\rho
^{\prime \prime \prime}}{4\rho^{\prime}}\right)  \rho \right.  \nonumber \\
&  \left.  +\frac{1}{2}\frac{d+1}{d+2}\rho^{\prime \prime}\right]  ,\nonumber \\
d= &  1,2,3.\label{eq72}%
\end{align}

%

\subsection{Explicit kinetic energy densities with $d=1,2,3$}

In the following we use the above results (\ref{mean_ked}) and (\ref{pos_ked_a}) with Eqs. (\ref{eq26}), (\ref{eq33}), (\ref{eq38}) and (\ref{a1}) to derive the expressions of the KED in $d=1,2,3$ dimensions with Airy functions. 

With $d=1$ we obtain
\begin{equation}
\tau^{d=1}(\xi)=g_{s}\frac{\hbar^{2}}{2m}\frac{1}{6l^{3}}\left[  2\xi^{2}%
A_{i}^{2}\left(  \xi \right)  -A_{i}\left(  \xi \right)  A_{i}^{\prime}\left(
\xi \right)  -2\xi A_{i}^{\prime2}\left(  \xi \right)  \right]  . \label{eq54}%
\end{equation}
And the expression of the positively defined KED in the gradient form is given by
\begin{equation}
\tau_{G}^{d=1}(\xi)=g_{s}\frac{\hbar^{2}}{2m}\frac{1}{3l^{3}}\left[  \xi
^{2}A_{i}^{2}\left(  \xi \right)  -2A_{i}\left(  \xi \right)  A_{i}^{\prime
}\left(  \xi \right)  -\xi A_{i}^{\prime2}\left(  \xi \right)  \right]  .
\label{eq56}%
\end{equation}

It should be noted that in Ref.~\cite{Brack-koch}, an
explicit analytical result for the KED was obtained for $d=1$ 
linear potential through the use of occupied single-particle states up to the Fermi energy. 
Our result in Eq.(\ref{eq56}) for the mean KED -- after including a
factor two for the spin degeneracy -- is similar to the expression given in
Eq. (A.7) of Ref.~\cite{Brack-koch}. 
This reference also includes an
expression for the Laplacian KED, but not for the positive KED. Since the latter is an important quantity in DFT, this KED is explicitly given above in Eq.~(\ref{eq56}). 

When $d=2$, the mean KED becomes
\begin{align}
\tau^{d=2}(\xi)  &  =g_{s}\frac{\hbar^{2}}{2m}\frac{2^{2/3}}{32\pi l^{4}%
}\left[  A_{i}\left(  2^{2/3}\xi \right)  +2^{2/3}\xi A_{i}^{\prime}\left(
2^{2/3}\xi \right)  \right. \nonumber \\
&  \left.  +2^{4/3}\xi^{2}A_{i1}\left(  2^{2/3}\xi \right)  \right],  
\label{eq59}%
\end{align}
and the positive KED in Eq. (\ref{eq43})
can be written as \bigskip%
\begin{align}
\tau_{G}^{d=2}(\xi)  &  =g_{s}\frac{\hbar^{2}}{2m}\frac{2^{2/3}}{32\pi l^{4}%
}\left[  3A_{i}\left(  2^{2/3}\xi \right)  +2^{2/3}\xi A_{i}^{\prime}\left(
2^{2/3}\xi \right)  \right. \nonumber \\
&  \left.  +2^{4/3}\xi^{2}A_{i1}\left(  2^{2/3}\xi \right)  \right]  .
\label{eq61}%
\end{align}
This expression is one of our key results.

Finally, when $d=3$ the KED can be written as
\begin{align}
\tau^{d=3}(\xi)=g_{s} &  \frac{\hbar^{2}}{2m}\frac{1}{20\pi l^{5}}\bigg[\left(
\frac{3}{4}-2\xi^{3}\right)  A_{i}^{2}\left(  \xi \right)  \nonumber \\
&  +\xi A_{i}\left(  \xi \right)  A_{i}^{\prime}\left(  \xi \right)  +2\xi
^{2}A_{i}^{\prime2}\left(  \xi \right)  \bigg],\label{eq63}%
\end{align}
and the positive KED becomes
\begin{align}
\tau_{G}^{d=3}(\xi)=g_{s} &  \frac{\hbar^{2}}{2m}\frac{1}{20\pi l^{5}}\left[
2\left(  1-\xi^{3}\right)  A_{i}^{2}\left(  \xi \right)  +\xi A_{i}\left(
\xi \right)  A_{i}^{\prime}\left(  \xi \right)  \right.  \nonumber \\
&  \left.  +2\xi^{2}A_{i}^{\prime2}\left(  \xi \right)  \right]  .\label{eq66}%
\end{align}
With $g_{s}=2$ this result is identical to the one obtained 
in Ref.~\cite{Lind-Matt-Armiento}. We can also examine the
density-functional form according to Eq.~(\ref{eq72}), which 
with $d=3$ becomes
\begin{equation}
\tau_{G}^{d=3}=\frac{\hbar^{2}}{2ml^{2}}\left[  -\frac{3}{5}\left(  \frac
{3}{2}\frac{\rho}{\rho^{\prime}}+\frac{\rho^{\prime \prime \prime}}%
{4\rho^{\prime}}\right)  \rho+\frac{2}{5}\rho^{\prime \prime}\right].
\label{eq73}%
\end{equation}
This expression can be compared to the result
obtained by Vitos~\cite{vitos2}. With
the present notation, that result reads
\begin{equation}
\tau_{\mathrm{Vitos}}^{d=3}=\frac{\hbar^{2}}{2ml^{2}}\left[  \frac{3}%
{5}\left(  \frac{\rho^{\prime \prime \prime}}{4\rho^{\prime \prime}}\frac
{3\rho \rho^{\prime \prime \prime}-2\rho^{\prime}\rho^{\prime \prime}}%
{2\rho^{\prime2}-3\rho \rho^{\prime \prime}}\right)  \rho+\frac{2}{5}%
\rho^{\prime \prime}\right]  . \label{eq74}%
\end{equation}
Although the above two expressions look very different, they are
actually equivalent. This is shown in detail in Appendix C. However, our analytical
expression in Eq. (\ref{eq73}) is much simpler to handle mathematically and
numerically than the one given in Eq. (\ref{eq74}).

\subsection{Refinement factor}

Here we derive a general expression for the so-called
refinement factor within the framework of Airy gas model in $d$ dimensions. In
a pioneering work by Baltin~\cite{baltin_naturforsch} an explicit KED expression, based on the Wigner-Kirkwood expansion~\cite{brack_bhaduri} and the linear potential approximation, was
obtained as
\begin{equation}
\tau_{G}^{d=3}=\tau_{\mathrm{TF}}^{d=3}\left[  \rho \right]  \widetilde
{\varkappa}_{Baltin}\left(  \left \vert \nabla\rho \right \vert
^{\frac{1}{2}}\rho^{-\frac{2}{3}}\right)  . \label{eq75}%
\end{equation}
Here $\tau_{\mathrm{TF}}^{d=3}\left[  \rho \right]  $ stands for the TF KED functional in $d=3$ dimensions given by
\begin{equation}
\tau_{\mathrm{TF}}^{d=3}\left[  \rho \right]  =\frac{\hbar^{2}}{2m}\frac{3}%
{5}(3\pi^{2})^{\frac{2}{3}}\text{ }\rho^{\frac{5}{3}}, \label{eq76}%
\end{equation}
and $\widetilde{\varkappa}_{Baltin}$ is a function of the scaled quantity
$|\nabla\rho|^{1/2}\rho^{-2/3}$. This function is called the {\em kinetic energy refinement factor}. Vitos et al. have examined the above
relation in the context of the Airy gas model~\cite{vitos2}. By leaving out
the Laplacian term (which vanishes upon the integration for any confined
system), the Airy gas KED expression can be 
written similarly to Eq. (\ref{eq75}), that is
\begin{equation}
\tau_{G}^{d=3}=\tau_{\mathrm{TF}}^{d=3}\left[  \rho \right]  \widetilde
{\varkappa}_{\rm Vitos}\left(  \xi \right)  \label{eq77}%
\end{equation}
with a refinement factor
\begin{equation}
\widetilde{\varkappa}_{\rm Vitos}\left(  \xi \right)  =-\frac{\xi}{l^{2}(3\pi
^{2})^{\frac{2}{3}}\text{ }\rho^{\frac{2}{3}}}. \label{eq78}%
\end{equation}
Here we have added a factor $l^{2}$ that is missing in Eq. (17) of Ref.~\cite{vitos2}. In that work, numerical studies show improvements brought by
Eq. (\ref{eq77}) compared to Eq. (\ref{eq75}). We point out that the above relation is
exact, and gradient corrections from the Airy gas, which are embedded in the
refinement factor, have been examined in Ref. \cite{Lind-Matt-Armiento}.

To proceed with a generalization of Eqs. (\ref{eq77})-(\ref{eq78}) to $d$
dimensions, we return to the examination of Eq. (\ref{eq72}). Here we use the
TF KED functional \cite{Brack2003,Bencheikh2005} given by
\begin{equation}
\tau_{\mathrm{TF}}^{(d)}\left[  \rho \right]  =\frac{\hbar^{2}}{2m}4\pi \left[
\frac{d}{4}\Gamma \left(  \frac{d}{2}\right)  \right]  ^{\frac{2}{d}}\frac
{d}{d+2}\text{ }\rho^{1+\frac{2}{d}}. \label{eq79}%
\end{equation}
It should be noted that at the TF level the three forms of the KED defined previously are identical. It is possible to recast the gradient
form in Eq.~(\ref{eq72}) as follows:
\begin{equation}
\tau_{G}^{(d)}(\xi)=\tau_{\mathrm{TF}}^{d}\left[  \rho \right]  \widetilde
{\varkappa}_{d}\left(  \xi \right)  \label{eq80}%
\end{equation}
with
\begin{equation}
\widetilde{\varkappa}_{d}\left(  \xi \right)  =-\frac{1}{4\pi l^{2}}\left(
\frac{4}{d\  \Gamma \left(  \frac{d}{2}\right)  }\right)  ^{\frac{2}{d}}%
\frac{\xi}{\text{ }\rho^{\frac{2}{d}}}. \label{eq81}%
\end{equation}
This expression constitutes a generalization of Eq. (\ref{eq78}). It is straightforward
to confirm that for $d=3$ Eq. (\ref{eq81}) reduces to Eq. (\ref{eq78}).


\section{Local virial theorem}

Let us consider a system of noninteracting fermions moving in a potential
$V(x)$. In the early work of March and Young~\cite{March-Young}, the so-called
differential virial theorem was derived: $\frac{\partial \tau(x)}{\partial
x}=-\frac{1}{2}\frac{\partial V(x)}{\partial x}\rho(x)$. This relation is a
version of the \emph{local} virial theorem, when the particle motion is restricted to one dimension ($d=1$). In general, a local virial theorem 
couples, at a given point $\bar{r}$ in space, the particle density, potential energy
and KED. The theorem has been generalised for the specific cases of an isotropic harmonic oscillator~\cite{Brack-koch} and a linear potential~\cite{Brack-koch,Bencheikh-Nieto} in $d$ dimensions.

Let us return to Eq. (\ref{eq49a}) and take the partial derivative of both
sides with respect to $z$, leading to
\begin{equation}
\frac{\partial \tau^{(d)}(z)}{\partial z}=-\frac{d}{2}F\left(  \frac{m}{2\pi
\hbar^{2}}\right)  ^{d/2}%
{\displaystyle \int \limits_{c-i\infty}^{c+i\infty}}
\frac{d\eta}{2\pi i}\frac{e^{\frac{\hbar^{2}}{24m}\eta^{3}F^{2}-\eta Fz}}%
{\eta^{1+d/2}}. \label{eq67}%
\end{equation}
 
Combining Eqs. (\ref{eq67}) and (\ref{eq18}) leads to a relationship
\begin{equation}
\frac{\partial \tau^{(d)}(z)}{\partial z}=-\frac{d}{2}\frac{\partial
v_{\mathrm{eff}}(z)}{\partial z}\rho^{(d)}(z),\label{eq69}%
\end{equation}
where $v_{\mathrm{eff}}(z)=Fz$. Hence, the local virial theorem holds for the
Airy gas model in $d$ dimensions.


\section{Local-density approximation}

Lieb and Simon~\cite{lieb_simon1, lieb_simon2} have proved that the TF
theory becomes exact in the limit $N\rightarrow \infty$. This
universal behavior in the bulk, together with universality near the
edge, have recently been examined at zero and nonzero
temperatures for a system of $N$ noninteracting fermions in a wide variety of
potentials~\cite{dean_doussal_majumdar_schehr}. Here show that for the Airy gas
model and well inside the bulk region, the KED becomes the TF KED functional 
in $d$ dimensions.

As $l$ measures the thickness of the edge region, we have $\xi=\frac{z}{l}\ll-1$
in the bulk, so that $\left \vert \xi \right \vert \gg1$%
~\cite{KohnMattson,Lind-Matt-Armiento}.
The LDA version of the positive KED in Eq. (\ref{eq72}) reads
\begin{equation}
\tau_{\mathrm{LDA}}^{(d)}\left[  \rho \right]  \approx-\frac{\hbar^{2}}{2ml^{2}%
}\frac{d}{d+2}\frac{d}{2}\frac{\rho^{2}}{\rho^{\prime}},\label{eq82}%
\end{equation}
where we have omitted the terms with derivatives higher than two. In this approximation Eq. (\ref{eq71a}) becomes
\begin{equation}
\xi \approx \frac{d}{2}\frac{\rho}{\rho^{\prime}},\label{eq83}%
\end{equation}
and writing this relation as $\xi^{-1}=2\rho^{\prime}/(d\rho)$, we obtain by
integration
\begin{equation}
\left \vert \xi \right \vert \approx C_{d}\text{ }\rho^{\frac{2}{d}},\label{eq84}%
\end{equation}
where $C_{d}$ is a positive constant determined below. Since in the considered
region we have $\xi \leqslant0$, so that $\xi=-\left \vert \xi \right \vert $, we can use Eq. (\ref{eq84}) to express the KED functional in Eq. (\ref{eq82}) as
\begin{equation}
\tau_{\mathrm{LDA}}^{(d)}\left[  \rho \right]  \approx+\frac{\hbar^{2}}{2ml^{2}%
}\frac{d}{d+2}\rho \left \vert \xi \right \vert .\label{eq85}%
\end{equation}
By substituting Eq. (\ref{eq84}) into (\ref{eq85}) we obtain the KED
functional
\begin{equation}
\tau_{\mathrm{LDA}}^{(d)}\left[  \rho \right]  \approx+\frac{\hbar^{2}}{2m}%
\frac{C_{d}\text{ }}{l^{2}}\frac{d}{d+2}\text{ }\rho^{1+\frac{2}{d}%
}.\label{eq86}%
\end{equation}
It is interesting to note that our expression in Eq. (\ref{eq86}) already
yields the correct density dependence, i.e., $\rho^{1+\frac{2}{d}}$, given by
the TF KED functional in $d$ dimensions [see Eq. (\ref{eq79})]. It remains
now to find the coefficient $C_{d}$ in Eq. (\ref{eq87}). Here we use
the explicit expressions of the particle density derived in Sec. II for
$d=1,2,3$. Furthermore, we can use the asymptotic
expressions for $\left \vert \xi \right \vert \gg1$ obtained from
Eqs. (10.4.60) and (10.4.62) in 
Ref.~\cite{Abramowitz-Stegun} with 
 the substitutions $z\rightarrow \left \vert \xi \right \vert $ and
$\zeta=$\textbf{ }$\frac{2}{3}z^{3/2}\rightarrow \frac{2}{3}\left \vert
\xi \right \vert ^{3/2}$. Thus, in the leading order we get
\begin{align}
A_{i}(-\left \vert \xi \right \vert )& \approx \frac{1}{\sqrt{\pi}\left \vert
\xi \right \vert ^{\frac{1}{4}}}\cos \left(  \frac{2}{3}\left \vert \xi \right \vert
^{\frac{3}{2}}-\frac{\pi}{4}\right),
\\
A_{i}^{\prime}(-\left \vert
\xi \right \vert )& \approx \frac{\left \vert \xi \right \vert ^{\frac{1}{4}}}%
{\sqrt{\pi}}\sin \left(  \frac{2}{3}\left \vert \xi \right \vert ^{\frac{3}{2}%
}-\frac{\pi}{4}\right)  .\label{eq87}%
\end{align}

Let us know examine the densities in $d$ dimensions.
According to Eq. (\ref{eq38}) the density 
with $d=1$ now becomes
\begin{equation}
\rho^{d=1}\approx \frac{2}{l\pi}\left \vert \xi \right \vert ^{\frac{1}{2}%
},\label{eq88}%
\end{equation}
where the factor two accounts for the spin degeneracy. We can rewrite Eq.
(\ref{eq88}) as $\left \vert \xi \right \vert \approx \pi^{2}l^{2}\left(
\rho^{d=1}\right)  ^{2}/4$. Upon comparing with Eq. (\ref{eq84}) for $d=1$, we
immediately find
\begin{equation}
C_{1}=\frac{\pi^{2}l^{2}}{4}.\label{eq89}%
\end{equation}

When $d=2$ we use the asymptotics of the primitive of Airy functions. To the
leading order we have, $A_{i1}\left(  -\left \vert t\right \vert \right)
\approx1$ for $\left \vert t\right \vert \gg1$ \cite{Vallee-Soares}. When
retaining only the leading order term, the density in Eq. (\ref{eq33}) reduces to
\begin{equation}
\rho^{d=2}\approx \frac{2}{4\pi l^{2}}\left \vert \xi \right \vert .\label{eq90}%
\end{equation}
We can write $\left \vert \xi \right \vert \approx2\pi l^{2}\rho^{d=2}$ and with
Eq. (\ref{eq84}) we obtain
\begin{equation}
C_{2}=2\pi l^{2}.\label{eq91}%
\end{equation}

In a similar way we first note that as $d=3$ the density in
Eq. (\ref{eq26}) reduces in the interior region to
\begin{equation}
\rho^{d=3}\approx \frac{1}{3\pi^{2}l^{3}}\left \vert \xi \right \vert ^{\frac{3}%
{2}}.\label{eq92}%
\end{equation}
Now we find $\left \vert \xi \right \vert \approx(3\pi^{2})^{\frac{2}{3}}%
l^{2}\left(  \rho^{d=3}\right)  ^{\frac{2}{3}}$. And using Eq. (\ref{eq84}) leads to
\begin{equation}
C_{3}=(3\pi^{2})^{\frac{2}{3}}l^{2}.\label{eq93}
\end{equation}
We can now express the above results for $C_{1}$, $C_{2}$ and $C_{3}$ in a
$d$-dependent form as
\begin{equation}
C_{d}=4\pi \left[  \frac{d}{4}\Gamma \left(  \frac{d}{2}\right)  \right]
^{\frac{2}{d}}l^{2}.\label{eq94}%
\end{equation}
Upon inserting this last expression into Eq.~(\ref{eq86}), we find a KED functional that is identical to that in Eq.~(\ref{eq79}). Hence, the KED of the Airy gas inside the bulk reduces to that of
the TF model, or to that of the LDA. An interesting extension of the present study would be going beyond the LDA limit and to find the
semiclassical Weizs\"{a}cker term of the KED given in $d$ dimensions by
$\left(  1-2/d\right)  (\nabla\rho)^{2}/12\rho$
\cite{Bencheikh2005}.


\section{Summary and outlook}

To summarize, we have used the widely studied Airy gas model to derive
explicit expressions for the edge particle density and for the corresponding
edge kinetic energy density (KED) in one, two, and three dimensions. 
Then we have obtained an expression for the positively defined KED in terms of the particle
density and its gradients in $d$ dimensions and shown that the
local virial theorem is satisfied. Finally, we have analyzed the
limit of the local-density approximation of the Airy gas model. We have shown
that in this limit the KED functional reduces to that of the Thomas-Fermi
model in $d$ dimensions. In a similar way as was suggested for the KED in
relation with the refinement factor, we believe that our findings in two and
one dimensions may be used for the exchange energy density in reduced
dimensions. In particular, our expressions for the density and KED may serve as inputs
in the expressions of exchange or exchange-correlation density functionals
developed in recent years for two-dimensional systems~\cite{vilhena,guandalini}.

\begin{acknowledgments}
This work has been supported by the Directorate General for Scientific
Research and Technological Development (DGRSDT) Algeria.
\end{acknowledgments}

\bigskip

\appendix

\section{Properties of Airy functions}
\label{appendixA}
Here we utilize the recent progress in the calculation of integrals
involving Airy functions as presented in Refs. \cite{Vallee-Soares} and
\cite{Abramowitz-Stegun}. The Airy function is defined as the solution to 
the following differential equation:
\begin{align}
A_{i}^{\prime\prime}(u)-u A_{i}(u)=0.
\label{a1}
\end{align}
Next we use the equation (3.86) in Ref.~\cite{Vallee-Soares}, i.e.,
\begin{align}
\int_{0}^{\infty}v^{-1/2}A_{i}\left(
u+v\right)  dv= 2^{2/3}\pi A_{i}^{2}\left(  \frac{u}{2^{2/3}}\right),\label{a2}
\end{align}
which leads to useful identities. The second derivative with respect to $u$ leads to~\cite{Abramowitz-Stegun}
\begin{align}
{\displaystyle \int \limits_{0}^{\infty}}
v^{1/2}  &  A_{i}\left(  u+v\right)  dv=\nonumber \\
&  2^{1/3}\pi \left[  A_{i}^{\prime2}\left(  2^{-2/3}u\right)  -2^{-2/3}%
uA_{i}^{2}\left(  2^{-2/3}u\right)  \right].  \label{a3}
\end{align}
The fourth derivative leads to~\cite{Abramowitz-Stegun}
\begin{align}
{\displaystyle \int \limits_{0}^{\infty}}
v^{3/2}  &  A_{i}\left(  u+v\right)  dv=\pi \bigg[\frac{u^{2}}{2^{1/3}}%
A_{i}^{2}\left(  \frac{u}{2^{2/3}}\right) \nonumber \\
-  &  A_{i}\left(  \frac{u}{2^{2/3}}\right)  A_{i}^{\prime}\left(  \frac
{u}{2^{2/3}}\right)  -2^{1/3}uA_{i}^{\prime2}\left(  \frac{u}{2^{2/3}}\right)
\bigg]. \label{a4}
\end{align}
And finally, the sixth derivative leads to
\begin{align}
{\displaystyle \int \limits_{0}^{\infty}}
v^{5/2} &  A_{i}\left(  u+v\right)  dv=\pi \left[  2^{-1/3}\left(  \frac{3}%
{2}-u^{3}\right)  A_{i}^{2}\left(  \frac{u}{2^{2/3}}\right)  \right.
\nonumber \\
&  \left.  +uA_{i}\left(  \frac{u}{2^{2/3}}\right)  A_{i}^{\prime}\left(
\frac{u}{2^{2/3}}\right)  +2^{1/3}u^{2}A_{i}^{\prime2}\left(  \frac{u}%
{2^{2/3}}\right)  \right].
\label{a5}
\end{align}
Changing the variables and using Eqs. (\ref{eq32}) and (\ref{a1}) leads to a useful
identity for the $d=2$ case: 
\begin{align}
{\displaystyle \int \limits_{0}^{\infty}}
v  &  A_{i}\left(  2^{2/3}\xi+v\right)  dv=\nonumber \\
&  -\left[  A_{i}^{\prime}\left(  2^{2/3}\xi \right)  +2^{2/3}\xi A_{i1}\left(
2^{2/3}\xi \right)  \right].  \label{a6}%
\end{align}


\section{Proof of Eq.~(\ref{eq71a})}
\label{appendixB}
From Eq. (\ref{der2rho})  we obtain the third derivative as
\begin{align}
\frac{\partial^{3}\rho(\xi)}{\partial \xi^{3}}
&= 4\left(\rho(\xi) + \xi\rho^{\prime}(\xi)\right)\nonumber\\
& + 2^{4/3} D_{d} 2^{2/3}
{\displaystyle \int \limits_{0}^{\infty}}
dv v^{1+d/2}A^{\prime}_{i}\left(  2^{2/3}\xi+v\right).\label{B1}
\end{align}
Since the last term can be integrated by parts and $A_{i}(\infty)=0$, we obtain 
\begin{align}
\frac{\partial^{3}\rho(\xi)}{\partial \xi^{3}}
&= 4\rho(\xi) + 4\xi\rho^{\prime}(\xi)\nonumber\\
&+ 4 D_{d}\left(- \frac{d+2}{2}\right)D^{-1}_{d}\rho \nonumber\\
&= 4\xi \rho^{\prime}-2d \rho. \label{B2}
\end{align}
This expression can be written in the form given in Eq. (\ref{eq71a}).


\section{Equivalence of Eqs. (\ref{eq73}) and (\ref{eq74})}
\label{appendixC}

To prove the equivalence of Eqs. (\ref{eq73}) and (\ref{eq74}), we 
compute the right-hand side of Eq. (\ref{eq74}) by
substituting the following explicit expressions for the ($d=3$)
density $\rho$ and its derivatives $\rho^{\prime}%
,\rho^{\prime \prime}$ and $\rho^{\prime \prime \prime}$.

In the $d=3$ case, let us rewrite Eq. (\ref{eq26}) as
\begin{equation}
12\pi l^{3}\rho=g_{s}\left[  2\xi(\xi A_{i}^{2}-A_{i}^{\prime2})-A_{i}%
A_{i}^{\prime}\right]  ,\tag{C1}%
\end{equation}
and recall that $A_{i}^{\prime \prime}(\xi)=\xi A(\xi)$. We deduce
\begin{align}
4\pi l^{3}\rho^{\prime} &  =g_{s}\left(  \xi A_{i}^{2}-A_{i}^{\prime2}\right)
\tag{C2}\\
4\pi l^{3}\rho^{\prime \prime} &  =g_{s}A_{i}^{2}\tag{C3}\\
2\pi l^{3}\rho^{\prime \prime \prime} &  =g_{s}A_{i}A_{i}^{\prime}.\tag{C4}%
\end{align}

Let us now return to Eq. (\ref{eq74}) and rewrite the term between the brackets in the right-hand
side as follows:
\begin{equation}
G=Q\times \frac{S}{K}\tag{C5}%
\end{equation}
with
\begin{align}
Q &  =\frac{\rho^{\prime \prime \prime}}{4\rho^{\prime \prime}},\tag{C6}\\
S &  =3\rho \rho^{\prime \prime \prime}-2\rho^{\prime}\rho^{\prime \prime},
\tag{C7}\\
K &  =2\rho^{\prime2}-3\rho \rho^{\prime \prime}.\tag{C8}%
\end{align}
By substituting Eqs. (C3) and (C4) into Eq. (C7) we get
\begin{equation}
Q=\frac{A_{i}^{\prime}}{2A_{i}}.\tag{C9}%
\end{equation}
Using Eqs. (C1-C4), Eq. (C7) becomes after simplifications%
\begin{equation}
S=\frac{A_{i}}{8\pi^{2}l^{6}}\left(  2\xi^{2}A_{i}^{2}A_{i}^{\prime}-2\xi
A_{i}^{\prime3}-\xi A_{i}^{3}\right)  .\tag{C10}%
\end{equation}
Similarly, we substitute Eqs. (C1-C4) into Eq. (C8) and find%
\begin{equation}
K=\frac{A_{i}^{\prime}}{16\pi^{2}l^{6}}\left(  2A_{i}^{\prime3}+A_{i}^{3}-2\xi
A_{i}^{2}A_{i}^{\prime}\right).  \tag{C11}%
\end{equation}
Upon insertion of these results into Eq. (C5) we find
\begin{align}
G &  =\frac{2\xi^{2}A_{i}^{2}A_{i}^{\prime}-2\xi A_{i}^{\prime3}-\xi A_{i}%
^{3}}{2A_{i}^{\prime3}+A_{i}^{3}-2\xi A_{i}^{2}A_{i}^{\prime}}\nonumber \\
&  =-\xi. \tag{C12}%
\end{align}
If we substitute this result into  Eq. (\ref{eq74}) the resulting
expression becomes identical to our result in Eq. (\ref{eq73}), since the quantity $\xi$ according to Eq. (\ref{eq71a}) with $d=3$ reduces
to $\xi=\left(  \frac{3}{2}\frac{\rho}{\rho^{\prime}}+\frac{\rho
^{\prime \prime \prime}}{4\rho^{\prime}}\right)$. Hence, we have shown
the equivalence of Eqs. (\ref{eq73}) and (\ref{eq74}).



\end{document}